\documentclass[10pt]{article}

\usepackage{amsmath,amssymb,amsthm}
\usepackage{amsxtra}
\usepackage{amsfonts}
\usepackage{amssymb}
\usepackage{url}
\usepackage{hyperref}
\usepackage{esint}
\usepackage{eucal}
\usepackage{mathrsfs}
\usepackage{cite}
\usepackage{graphicx,graphics}
\newcommand\fig[1] {{\rm Figure}~\ref{fig:#1}}
\newcommand\labfig[1] {\label{fig:#1}}
\newcommand\sect[1] {\ref{sect:#1}}
\newcommand\labsect[1] {\label{sect:#1}}

\newcommand\eq[1] {(\ref{#1})}
\newcommand{\bfm}[1]{\mbox{\boldmath ${#1}$}}

\newcommand{\nonum}{\nonumber \\}
\newcommand{\beqa}{\begin{eqnarray}}
\newcommand{\eeqa}[1]{\label{#1}\end{eqnarray}}
\newcommand{\beq}{\begin{equation}}
\newcommand{\eeq}[1]{\label{#1}\end{equation}}
\newcommand{\Grad}{\nabla}
\newcommand{\Div}{\nabla \cdot}
\newcommand{\Curl}{\nabla \times}

\newcommand{\Real}{\mathop{\rm Re}\nolimits}
\newcommand{\Imag}{\mathop{\rm Im}\nolimits}
\newcommand{\Tr}{\mathop{\rm Tr}\nolimits}

\newcommand{\und}[1]{\smash{\underline{#1}}}

\newcommand{\Ga}{\alpha}
\newcommand{\Gb}{\beta}

\newcommand{\Ge}{\epsilon}

\newcommand{\Gf}{\phi}

\newcommand{\Gg}{\gamma}
\newcommand{\Gc}{\chi}

\newcommand{\Gl}{\lambda}

\newcommand{\Gv}{\nu}

\newcommand{\Gvt}{\vartheta}

\newcommand{\Go}{\omega}

\newcommand{\GO}{\Omega}

\newcommand{\BGve}{\bfm\varepsilon}

\newcommand{\BGn}{\bfm\eta}
\newcommand{\BGm}{\bfm\mu}

\newcommand{\BGs}{\bfm\sigma}

\newcommand{\BGG}{\bfm\Gamma}

\newcommand{\CA}{{\cal A}}

\newcommand{\CE}{{\cal E}}

\newcommand{\CJ}{{\cal J}}

\newcommand{\CS}{{\cal S}}
\newcommand{\CT}{{\cal T}}
\newcommand{\CU}{{\cal U}}

\newcommand{\BCC}{{\bfm{\cal C}}}

\newcommand{\BCT}{{\bfm{\cal T}}}

\newcommand{\bpm}{\begin{pmatrix}}
\newcommand{\epm}{\end{pmatrix}}

\def\Be{{\bf e}}

\def\Bj{{\bf j}}
\def\Bk{{\bf k}}

\def\Bs{{\bf s}}

\def\Bu{{\bf u}}
\def\Bv{{\bf v}}

\def\Bx{{\bf x}}

\def\Bz{{\bf z}}
\def\BA{{\bf A}}
\def\BB{{\bf B}}

\def\BE{{\bf E}}
\def\BF{{\bf F}}
\def\BG{{\bf G}}
\def\BH{{\bf H}}
\def\BI{{\bf I}}
\def\BJ{{\bf J}}

\def\BL{{\bf L}}
\def\BM{{\bf M}}
 
\def\BO{{\bf O}}
\def\BP{{\bf P}}
\def\BQ{{\bf Q}}
\def\BR{{\bf R}}

\def\BT{{\bf T}}
\def\BU{{\bf U}}
\def\BV{{\bf V}}

\def\BZ{{\bf Z}}

\usepackage{graphics}
\usepackage{amssymb}

\title{
    A unifying perspective on linear continuum equations prevalent in science. Part V: resolvents; bounds on their spectrum; and their Stieltjes integral representations when the operator is not selfadjoint.}
\author{}
\date{}
\begin{document}
\maketitle
\vskip -.5cm
\centerline{\large Graeme W. Milton}
\centerline{Department of Mathematics, University of Utah, USA -- milton@math.utah.edu.}
\vskip 1.cm
\begin{abstract}
  We consider resolvents of operators taking the form $\BA=\BGG_1\BB\BGG_1$
  where $\BGG_1(\Bk)$ is a projection that acts locally in Fourier space and $\BB(\Bx)$ is an
  operator that acts locally in real space. Such resolvents arise naturally when one wants
  to solve any of the large class of linear physical equations
  surveyed in Parts I, II, III, and IV 
  that can be reformulated as problems in the extended abstract theory of composites.
  We review how $Q^*$-convex operators can be used to bound the spectrum of $\BA$.  Then, based on the Cherkaev-Gibiansky transformation and subsequent developments, that we reformulate, we obtain for non-Hermitian $\BB$, a 
  Stieltjes type integral
  representation for the resolvent $(z_0\BI-\BA)^{-1}$. The representation holds
  in the half plane  $\Real(e^{i\vartheta}z_0)>c$,
  where $\vartheta$ and $c$ are such that $c-[e^{i\vartheta}\BB+e^{-i\vartheta}\BB^\dagger]$ is positive definite (and coercive).
  \end{abstract}
%%%%%%%%%%%%%%%%%%%%%%%%%%%%%%%%%%%%%%%%%%%%%%%%%%%%%%%%%%%%%%%%%%%%%%%% 
\section{Introduction}
\setcounter{equation}{0}
\labsect{30}
%%%%%%%%%%%%%%%%%%%%%%%%%%%%%%%%%%%%%%%%%%%%%%%%%%%%%%%%%%%%%%%%%%%%%%%%%%%%%%%%%%%%%%%%%%%%%%%%%%%%%%%%%%%
In Parts I, II, III, and IV  \cite{Milton:2020:UPLI, Milton:2020:UPLII, Milton:2020:UPLIII, Milton:2020:UPLIV} we established that an avalanche of equations in science can be rewritten in the form
\beq \BJ(\Bx,t)=\BL(\Bx,t)\BE(\Bx,t)-\Bs(\Bx,t),\quad \BGG_1\BE=\BE,\quad\BGG_1\BJ=0,
\eeq{ad1}
as encountered in the extended abstract theory of composites, $\BGG_1$ is a projection
operator that acts locally in Fourier space, and $\Bs(\Bx)$ is the source term.

Here in Part V we are concerned with resolvents of the form
\beq \BR_0=(z_0\BI-\BA)^{-1}=z_0(\BI-\BA/z_0)^{-1}, \eeq{0.1}
(that, as we will see in the next section, arise naturally in the solution of \eq{ad1})
where the operator $\BA$ takes the form $\BA=\BGG_1\BB\BGG_1$, in which $\BGG_1$ is a projection operator in Fourier space, while
$\BB$ acts locally in real space and typically has an inverse,
and one that is easily computed. 
Thus if $\BGG_1$ or $\BB$ act on a field $\BF$ to produce a field $\BG$ then we have, respectively,
that $\BG(\Bx)=\BB(\Bx)\BF(\Bx)$ or $\widehat{\BG}(\Bk)=\BGG_1(\Bk)\widehat{\BF}(\Bk)$, in
which $\widehat{\BG}(\Bk)$ and $\widehat{\BF}(\Bk)$ are the Fourier components of $\BG$ and $\BF$.

As in the previous parts
we define the inner product of two fields $\BP_1(\Bx)$ and $\BP_2(\Bx)$ to be
\beq (\BP_1,\BP_2)=\int_{\mathbb{R}^3}(\BP_1(\Bx),\BP_2(\Bx))_{\CT}\,d\Bx,
\eeq{innp}
where $(\cdot,\cdot)_{\CT}$ is a suitable inner product on the space $\CT$
such that the projection $\BGG_1$ is selfadjoint with
respect to this inner product, and thus the space $\CE$ onto which
$\BGG_1$ projects is orthogonal to the space $\CJ$ onto which
$\BGG_2=\BI-\BGG_1$ projects. We define the norm of a field $\BP$ to
be $|\BP|=(\BP,\BP)^{1/2}$, and given any operator $\BO$ we define its norm
to be
\beq \|\BO\|=\sup_{\BP,|\BP|=1}|\BO\BP|. \eeq{on}
When we have periodic fields in periodic media the integral in \eq{innp} should be taken over the unit cell $\GO$ of periodicity.
If the fields depend on time $t$ then we should set $x_4=t$ take the integral over $\mathbb{R}^4$ with the integral over the spatial variables
restricted to $\GO$ if the material and fields are spatially periodic. 

The goal of this paper is four fold:
\begin{itemize}
\item To highlight the connection between such resolvents and the solution of
problems in the extended theory
of composites;

\item To review how $Q^*$-convex operators can be used to bound the
spectrum of $\BA=\BGG_1\BB\BGG_1$, and to review some methods for constructing
$Q^*$-convex operators \cite{Milton:2019:NRF};

\item To establish a remarkable connection, founded on the work of Cherkaev and Gibiansky \cite{Cherkaev:1994:VPC} and elaborated upon
  in \cite{Milton:1990:CSP, Milton:2009:MVP, Milton:2010:MVP, Milton:2017:BCP}, between the resolvent
  with $\BA=\BGG_1\BB\BGG_1$ where $\BA$ is not Hermitian, and
  the inverse $\BH^0$ of an associated operator having  Hermitian real and imaginary parts,
  and with $\BH^0$ being real and positive definite when $z_0$ is real and greater than a constant
  $c$ such that $c-[\BB+\BB^\dagger]$ is positive definite. Furthermore the Hermitian part
  of $\BH^0$ is positive definite in the half plane $\Real z_0>c$.

\item On the basis of this connection to obtain Stieltjes type
  integral representations for the resolvent in the case where $\BB$ is
  non-selfadjoint but there exists an angle $\vartheta$
  such that $c-[e^{i\vartheta}\BB+e^{-i\vartheta}\BB^\dagger]$ is positive definite
  (and coercive) for some constant $c$. The integral representation holds
  in the half plane $\Real(e^{i\vartheta}z_0)>c$.
  
\end{itemize}

\noindent The work presented is largely based on the articles \cite{Cherkaev:1994:VPC, Milton:1990:CSP,
  Milton:2009:MVP, Milton:2010:MVP, Milton:2019:NRF}, but develops some of the ideas further.

There is also the related resolvent
\beqa \BR & \equiv & (z_0\BI-\BGG_1\BB)^{-1}\BGG_1=(z_0\BGG_1-\BA)^{-1} \nonum
         & = & (z_0\BI-\BA)^{-1}\BGG_1=\BR_0+(\BGG_1-\BI)/z_0,
\eeqa{0.1a}
where in the final expression on the first line the inverse is to be taken on the subspace onto which  $\BGG_1$ projects -- thus $\BR$ is the resolvent of
$\BA$ within this subspace, i.e. on this subspace
\beq \BR=(\BGG_1\BL\BGG_1)^{-1}\quad \text{where}\quad \BL=z_0\BI-\BB.
\eeq{0.1aaa}
The equivalences in \eq{0.1a} are easily checked by expanding each expression in a powers of $\BA$ or $\BGG_1\BB$. 

One reason for the importance of knowing the resolvent as a function of $z$ is that it allows computation of any operator
valued analytic function $f(\BA)$ of the matrix $\BA$ according to
the formula
\beq f(\BA)=\frac{1}{2\pi i}\int_\Gg f(z_0)(z_0\BI-\BA)^{-1}\,dz_0,
\eeq{-0.1d}
where $\Gg$ is a closed contour in the complex plane that encloses the spectrum of $\BA$.

The first equation in \eq{ad1} is called the constitutive law with $\Bs(\Bx)$ being the source term. As remarked previously, if the null space of $\BL$ is nonzero
then one may one can often shift $\BL(\Bx)$ by a
multiple $c$ of a ``null-$\BT$ operator''
$\BT_{nl}(\Bx)$ (acting locally in real space or spacetime, and discussed further in Section \sect{34}), defined to have the
property that
\beq \BGG_1\BT_{nl}\BGG_1=0, \eeq{nl-1}
that then has an associated quadratic form (possibly zero) that is
a ``null-Lagrangian''.
Clearly the equations \eq{ad1} still hold, with $\BE(\Bx)$ unchanged
and $\BJ(\Bx)$ replaced by $\BJ(\Bx)+c\BT_{nl}\BE(\Bx)$
if we replace $\BL(\Bx)$ with  $\BL(\Bx)+c\BT_{nl}(\Bx)$. In other
cases $\BL$ may contain $\infty$ (or $\infty$'s) on its diagonal.
If one can remove any degeneracy of $\BL(\Bx)$, we can consider the
dual problem
\beq \BE=\BL^{-1}\BJ(\Bx)+\BL^{-1}\Bs(\Bx),\quad \BGG_2\BJ=\BJ,
\quad\BGG_2\BE=0,
\eeq{dual}
with $\BGG_2=\BI-\BGG_1$, and then, if desired, try to shift  $\BL^{-1}(\Bx)$
by a multiple
of a ``null-$\BT$ operator'' $\widetilde{\BT}_{nl}(\Bx)$ satisfying
$\BGG_2\widetilde{\BT}_{nl}\BGG_2=0$ to remove its degeneracy.

Our results, in particular, apply to the family of problems associated with analyzing the response of two phase composite materials, 
where $\BB(\Bx)$ itself depends on $z_0$ and takes the form
\beq \BB(\Bx)=z_0\BI-\BL_1\Gc_1(\Bx)-\BL_2\Gc_2(\Bx), \eeq{0.2}
where the $\Gc_i(\Bx)$ are the characteristic functions
\beqa\Gc_i(\Bx) &= & 1\quad \text{in phase}\,\,i \nonum
&= & 0\quad \text{elsewhere},
\eeqa{0.3}
satisfying $\Gc_1(\Bx)+\Gc_2(\Bx)=1$, while $\BL_1$ and $\BL_2$ are the tensors of the two phases, representing their material properties, and
the ``reference parameter'' $z_0$ can be freely chosen. In the particular case
when $\BL_1=z_1\BI$ and $\BL_2=z_2\BI$ we have
\beq \BB(\Bx)=z_0\BI-z_1\BI\Gc_1(\Bx)-z_2\BI\Gc_2(\Bx)=(z_0-z_2)\BI-(z_1-z_2)\BI\Gc_1(\Bx), \eeq{0.4}
where now, for example, $z_1$ and $z_2$ may represent the conductivities of the two phases and $z_0$ a reference conductivity. With the choice
$z_0=z_2$ the expression \eq{0.1} reduces to
\beq \BR=z_2^{-1}\{\BI-(1-z_1/z_2)\BGG_1\Gc_1\BGG_1\}^{-1},
\eeq{0.5}
which is now again a problem directly of the form \eq{0.1} with $\BB$ and $z_0$ now being identified as
\beq \BB=\Gc_1\BI,\quad z_0=z_2/(z_2-z_1). \eeq{0.5a}
%In the case where only $\BL_2=z_2\BI$ a similar analysis gives
%\beq \BR=z_2^{-1}\{\BI-(1-z_1/z_2)\BGG_1\Gc_1(\BI-z_1\BL_1/z_2)\BGG_1\}^{-1}.
%\eeq{0.5mod}

%%%%%%%%%%%%%%%%%%%%%%%%%%%%%%%%%%%%%%%%%%%%%%%%%%%%%%%%%%%%%%%%%%%%%%%%%%%%%%%%%%%%%%%%%%%%%%%%%%%%%%%%%%%

%%%%%%%%%%%%%%%%%%%%%%%%%%%%%%%%%%%%%%%%%%%%%%%%%%%%%%%%%%%%%%%%%%%%%%%%%%%%%%%%%%%%%%%%%%%%%%%%%%%%%%%%%%%
\section{Recasting the resolvent problem as a problem in the extended abstract theory of composites}
\setcounter{equation}{0}
\labsect{31}
%%%%%%%%%%%%%%%%%%%%%%%%%%%%%%%%%%%%%%%%%%%%%%%%%%%%%%%%%%%%%%%%%%%%%%%%%%%%%%%%%%%%%%%%%%%%%%%%%%%%%%%%%%%%%%%%%%%%
As $\BGG_1$ is a selfadjoint projection in Fourier space, so too is $\BGG_2=\BI-\BGG_1$. We let $\CE$ denote the space of fields onto
which $\BGG_1$ projects, and $\CJ$ denote the orthogonal space of fields onto
which $\BGG_2$ projects. Associated with these operators is a problem in the extended abstract theory of composites: given $\Bs\in\CE$, find
$\BE\in\CE$ and $\BJ\in\CJ$ that solve
\beq \BJ=\BL\BE-\Bs.
\eeq{1.3}
The abstract theory of composites is reviewed, for example,
in Chapter 12 and forward in \cite{Milton:2002:TOC},
and in Chapters 1 and 2 in \cite{Milton:2016:ETC}. The extended abstract theory of composites is developed in Chapter 7 of
\cite{Milton:2016:ETC} and further in \cite{Milton:2019:NRF}.
Applying $\BGG_1$ to both sides of \eq{1.3} we obtain 
\beq \BGG_1\BL\BE=\BGG_1\Bs, \eeq{1.4}
which has the solution
\beq \BE=[\BGG_1\BL\BGG_1]^{-1}\BGG_1\Bs. \eeq{1.5}

Next let us introduce a constant Hermitian (typically positive definite) reference tensor
$\BL_0$, the associated ``polarization field''
\beq \BP=\BJ-\BL_0\BE=(\BL-\BL_0)\BE-\Bs, \eeq{1.6}
and a matrix $\BGG$ defined by
\beq \BE'=\BGG\BP \text{  if and only if  } \BE'\in\CE,\quad \BP-\BL_0\BE'\in\CJ. \eeq{1.7}
Equivalently, $\BGG$ can be defined by
\beq \BGG=\BGG_1(\BGG_1\BL_0\BGG_1)^{-1}\BGG_1,
\eeq{1.8}
where the inverse is to be taken on the subspace $\CE$. The operator $\BL_0\BGG$ is the projection onto the subspace $\BL_0\CE$ that annihilates $\CJ$.
These are not orthogonal subspaces unless one modifies the norm to $(\BP_1,\BP_2)_{\BL_0}=(\BL_0\BP_1,\BP_2)$ \cite{Moulinec:2018:CIM} but this
brings with it the problem that if $\BL(\Bx)$ is Hermitian in the original norm, then it will not be in the
new norm unless $\BL(\Bx)$ commutes with $\BL_0$. Instead, if $\BL_0$ is positive definite
then $\BL_0^{1/2}\BGG\BL_0^{1/2}$ is the projection onto  $\BL_0^{1/2}\CE$
that annihilates the orthogonal subspace $\BL_0^{-1/2}\CJ$.
We require that $\BL_0$ be chosen so that the rank of $\BGG_1(\Bk)\BL_0\BGG_1(\Bk)$ does not change as $\Bk$ varies. While the operator
$\BGG$ is not a projection with respect to the standard norm it satisfies
\beq \BGG\BL_0\BGG=\BGG.
\eeq{1.13}

From \eq{1.6} it follows that $\BGG\BP=-\BE$, and so
\beq \BGG\BP=\BGG(\BL-\BL_0)\BE-\BGG\Bs=-\BE, \eeq{1.10}
implying
\beq \BE=[\BI-\BGG\BB]^{-1}\BGG\Bs,\text{  where } \BB\equiv\BL_0-\BL. \eeq{1.10a}
Comparing this with \eq{1.5} gives
\beq [\BGG_1\BL\BGG_1]^{-1}\BGG_1=[\BI-\BGG\BB]^{-1}\BGG. \eeq{1.11}
While it appears like the right hand side of \eq{1.11} depends
on $\BL_0$, and not just on $\BL$ and $\BGG_1$
the identity and the preceding derivation shows it does not.
In particular, a general choice of $\BL_0$ gives the same result as the choice $\BL_0=z_0\BI$,
for which $\BGG=\BGG_1/z_0$. This establishes the identity
\beq [\BI-\BGG\BB]^{-1}\BGG=[z_0\BI-\BGG_1\BB_0]^{-1}\BGG_1=\BR
\eeq{1.11id}
where $\BR$ is the resolvent \eq{0.1a} and $\BB_0=z_0\BI-\BL=\BB+z_0\BI-\BL_0$.
Conversely, if we are interested in computing the
resolvent in \eq{0.1} or \eq{0.1a}, then we can recast it as a problem in the theory of composites with $\BL=\BL_0-\BB$, where we are
free to choose $\BL_0$. The solution \eq{1.10a} is well known in the theory of composites: see, for example Chapter 14 of \cite{Milton:2002:TOC},
\cite{Willis:1981:VRM}, and references therein.

For the special case of a two phase medium where $\BB(\Bx)$ takes the
form \eq{0.2} we may take $\BL_0=\BL_2$ giving
$\BB(\Bx)=\Gc(\Bx)(\BL_2-\BL_1)$ and correspondingly
\beq \BR=[\BI-\BGG\Gc(\Bx)(\BL_2-\BL_1)]^{-1}\BGG.
\eeq{2ph}

Having established this connection with the resolvent we can now apply all the theory developed in extended abstract theory
of composites to resolvents of the required form, and conversely. In the theory of composites it is clear that \eq{1.3} can be written
in the equivalent form
\beq  \BE=\BL^{-1}\BJ+\BL^{-1}\Bs. \eeq{1.12a}
So direct analogy with \eq{1.10a} gives
\beq \BJ=-[\BI-\widetilde{\BGG}\widetilde{\BB}]^{-1}\widetilde{\BGG}\BL^{-1}\Bs, \eeq{1.13a}
where 
\beq \widetilde{\BB}=\BM_0-\BL^{-1}, \quad \widetilde{\BGG}=\BGG_2(\BGG_2\BM_0\BGG_2)^{-1}\BGG_2,
\eeq{1.14}
in which we can freely choose $\BM_0$.

Substituting \eq{1.13a} in \eq{1.12a} and comparing the result with \eq{1.5} gives
\beq [\BI-\BGG\BB]^{-1}\BGG=\BL^{-1}-\BL^{-1}[\BI-\widetilde{\BGG}\widetilde{\BB}]^{-1}\widetilde{\BGG}\BL^{-1},
\eeq{1.15}
and in particular with $\BM_0=\BI/z_0$ this implies
\beq  \BR=[z_0\BI-\BGG_1\BB]^{-1}\BGG_1=\BL^{-1}-\BL^{-1}\{\BI/z_0-\BGG_2[\BI/z_0-(z_0\BI-\BB)^{-1}]\}^{-1}\BGG_2\BL^{-1}.
\eeq{1.15a}
It follows from this identity that if $\BB(\Bx)=\chi(\Bx)\BI$ where $\chi(\Bx)$ is a characteristic function, then
\beq \BR=[z_0\BI-\BGG_1\BB]^{-1}\BGG_1=
\BL^{-1}-z_0(1-z_0)\BL^{-1}[(1-z_0)\BI-\BGG_2\BB]^{-1}\BGG_2\BL^{-1}.
\eeq{spectI}
So if $\Gl$ is in the spectrum of $\BGG_1\BB\BGG_1$ then $1-\Gl$ will be in the spectrum of $\BGG_2\BB\BGG_2$.
This can also be established by representing $\chi\BI$ in a basis where $\BGG_1$ is block diagonal with
$\BI$ in the first block and 0 in all other entries.

There is another connection with the theory of composites. Suppose that $\Bs\in\CE$ is given and define the three projection operators
\beq \overline{\BGG}_0=\Bs\otimes\Bs/|\Bs|^2,\quad \overline{\BGG}_1=\BGG_1-\overline{\BGG}_0,\quad
\overline{\BGG}_2=\BGG_2=\BI-\overline{\BGG}_0-\overline{\BGG}_1, \eeq{r.1}
and the three subspaces $\overline{\CU}$, $\overline{\CE}$, and $\CJ$ onto which they project. Now the solution
to $\BE$ to \eq{1.3} can be written as $\BE=\overline{\BE}+z_*\Bs$, where $\overline{\BE}\in\overline{\CE}$. Hence \eq{1.3}
can be recast as a standard problem in the abstract theory of composites: given $\Bs\in\CU$ find $z_*$, which is known in the theory of composites as an effective parameter, such that
\beq (\BJ+\Bs)=\BL(\overline{\BE}+\Bs/z_*),\quad \BJ\in\CJ,\quad \overline{\BE}\in \overline{\CE}, \eeq{r.2}
and the solution \eq{1.10a} implies
\beq \Bs/z_*=\BGG_0[z_0\BI-\BGG_1\BB]^{-1}\BGG_1\Bs,\text{  where } \BB\equiv z_0\BI-\BL. \eeq{r.3}
Now we treat $\BB$ as fixed so that $\BL$ depends on $z_0$.
Consider the function $z_*(z_0)$. When $z_0$ has positive (negative) imaginary part,
and $\BB$ is a matrix, the resolvent $[z_0\BI-\BGG_1\BB ]^{-1}$ is negative (positive) definite.
We conclude that the imaginary part of $z_*$ has the same sign as that of $z_0$.
When $\BB$ is a finite dimensional matrix the poles and zeros of $z_*$ lie on the real axis, are simple, and the poles alternate with the zeros along the real axis. Assuming the source $\Bs$ excites all modes (i.e is not orthogonal to any eigenfunction of $\BGG_1\BB \BGG_1$)
the zeros of $z_*(z_0)$ will reveal the spectrum of $[z_0\BI-\BGG_1\BB ]^{-1}$.

Owing to the invariance of the form of \eq{r.3} when we make the
replacements
\beq \Bs\to z_*\Bs, \quad \BJ\to\und{\BE},\quad \und{\BE}\to\BJ,\quad\BL\to\BL^{-1},
\eeq{r.4}
we obtain the alternative identity:
\beq z_*\Bs=\BGG_0[\und{z}_0-\BGG_2\und{\BB}]^{-1}\BGG_2\BGG_0\Bs,
\text{  where }\und{z}_0=1/z_0,\quad
\und{\BB}\equiv \und{z}_0\BI-\BL^{-1}=\BI/z_0-[\BI/z_0-\BB]^{-1}.
\eeq{r.2a}
Treating $\und{\BB}$ as fixed we see that the spectrum of
$[\und{z}_0-\BGG_2\und{\BB}\BGG_2]^{-1}$ is revealed by the spectrum
of $z_*(z_0)$. 
%%%%%%%%%%%%%%%%%%%%%%%%%%%%%%%%%%%%%%%%%%%%%%%%%%%%%%%%%%%%%%%%%%%%%%%%%%%%%%%%%%%%%%%%%
\section{Bounding the spectrum of $\BA$ and $Q^*$-convexity}
\setcounter{equation}{0}
\labsect{34}
%%%%%%%%%%%%%%%%%%%%%%%%%%%%%%%%%%%%%%%%%%%%%%%%%%%%%%%%%%%%%%%%%%%%%%%%%%%%%%%%%%%%%%%%%%%%%%%%%%%%%%%%%%%%%%%%%%%%
The spectrum of $\BA$ consists of those values of $z_0$ for which the inverse of
$z_0\BI-\BA$ does not exist. Let us assume that $\BB$ and hence $\BA$ are self adjoint.
Then the spectrum is on the real axis and we let $[\Ga^-,\Ga^+]$ denote the smallest interval
on the real axis that contains all the spectrum ($\Ga^-$ could be $-\infty$
and $\Ga^+$ could be $+\infty$). Here we interested in finding
outer bounds on the spectrum: constants $a^-$ and $a^+$ such that $[\Ga^-,\Ga^+]\subset[a^-,a^+]$; and outer bounds on the spectrum: constants $c^-$ and $c^+$ such that $[c^-,c^+]\subset[\Ga^-,\Ga^+]$.
Inner bounds on the spectrum allow one to see how tight are outer bounds,
and vice-versa. 

The most well known inner bounds on the spectrum are those obtained by the
Rayleigh-Ritz method: one may look for the extreme lower value $c^-_{RR}$
and extreme upper value
$c^+_{RR}$ of $(\BA\Bs,\Bs)/|\Bs|^2\in[\Ga^-,\Ga^+]$ as $\Bs$ varies in some
finite dimensional subspace $\CS$. Let  $\Bs=\Bs^-_{RR}$ and $\Bs=\Bs^+_{RR}$
be the corresponding fields, normalized with $|\Bs^-_{RR}|=|\Bs^+_{RR}|=1$,
that achieve these extreme values. The Rayleigh-Ritz method can be improved
by using the power method as 
$|((\BA-c\BI)^n\Bs,\Bs)|/|\Bs|^2$ provides a lower bound on the maximum
of $|\Ga^--c|^n$ and $|\Ga^+-c|^n$. This gives the bounds
\beq c^+=c^-_{RR}+|((\BA-c^-_{RR}\BI)^n\Bs^+_{RR},\Bs^+_{RR})|^{1/n}, \quad
c^-=c^+_{RR}-|((\BA-c^+_{RR}\BI)^n\Bs^-_{RR},\Bs^-_{RR})|^{1/n}
\eeq{ibs}
where they hold provided $\CS$ is large enough so that
$|\Ga^+-c^-_{RR}|^n$ exceeds $|c^-_{RR}-\Ga^-|^n$ in the first case
and $|c^+_{RR}-\Ga^-|^n$ exceeds $|\Ga^+-c^+_{RR}|^n$ in the second case.
Outer bounds on the spectrum, that we come to now, can be used to verify
that these conditions hold.

Outer bounds on the spectrum can be obtained using
the powerful methods introduced in \cite{Milton:2019:NRF} to bound resolvents using $Q^*$-convex operators. The methods build on a large body
of literature associated with quasiconvex functions and the associated notion of weak lower semicontinuity. This has a
long history, with many applications, reviewed for example in \cite{Dacorogna:1982:WCW,Benesova:2017:WLS}.
The approach can have the advantage that in one fell
swoop it gives bounds that are universally valid for the spectrum of all resolvents in a given class. For example, if $\BL(\Bx)$ is piecewise constant
taking $N$ values,
that then can be labeled as $N$ different phases, then the bounds on the spectrum can be independent of the geometry, i.e., on the way these phases are
distributed. As yet, due to their novelty, the methods described have not been exploited, even within the theory of composites. That they
will prove to be a strong tool is guaranteed, based on the successful application of quasiconvex functions for obtaining sharp
bounds on effective moduli based on the translation method, or method of compensated compactness, as summarized in
the books \cite{Cherkaev:2000:VMS, Torquato:2001:RHM, Milton:2002:TOC, Allaire:2002:SOH, Tartar:2009:GTH}. I prefer the name translation method
as it applies more broadly (see, for example, \cite{Milton:2013:SIG}) than within the  compensated compactness
framework of sequences of spatially oscillating fields with progressively
finer and finer oscillations (as occurs in periodic homogenization when one has a sequence of periodic materials with
smaller and smaller unit cell sizes) --- in particular the concept of
$Q^*$-convexity, described below, loses its significance in the
compensated compactness setting. In practice, if one is not considering optimization problems
or energy minimization problems, then one rarely has a sequence of materials, but rather just one inhomogeneous material and one wants to say something about the response of it.
For geometries of well separated spheres Bruno \cite{Bruno:1991:ECS} obtained some bounds on the spectrum for the conductivity problem.

Let us first consider the case when $\BB$ and hence $\BA$ are Hermitian. Then, to bound the spectrum of $\BA$ we
look for a  Hermitian tensor field $\BT(\Bx)$, and constant $a^-$ such that:
\beq \BB(\Bx)\geq \BT(\Bx) +a^-\BI\quad\text{for all }\Bx, \quad \BGG_1\BT\BGG_1\geq 0. \eeq{t.1}
Following \cite{Milton:2013:SIG, Milton:2019:NRF} we call $\BT(\Bx)$ a $Q^*$-convex operator and the associated quadratic form a $Q^*$-convex function.
$Q^*$-convexity generalizes, for quadratic forms, the notion of 
quasiconvex functions, which have a long history, reviewed for example in \cite{Dacorogna:1982:WCW,Benesova:2017:WLS}.
Besides their importance for obtaining bounds on the effective properties of
composites as outlined in the books \cite{Cherkaev:2000:VMS,Torquato:2002:RHM,Milton:2002:TOC,Allaire:2002:SOH,Tartar:2009:GTH,Milton:2016:ETC}, they have been a
powerful tool for the proof of existence of solutions to the nonlinear
Cauchy elasticity equations \cite{Ball:1977:CCE,Ball:1981:NLW}
and for furthering our understanding of shape memory material alloys \cite{Ball:1987:FPM,Ball:1992:PET}.

For quadratic functions, quasiconvexity \cite{Morrey:1952:QSM, Morrey:1966:MICBook, Meyers:1965:ASA}, and the closely related $\CA$-quasiconvexity \cite{Fonseca:2006:AQL}
are associated with $Q^*$-convexity when $\BT(\Bx)$ is independent of $\Bx$ and
$\BGG_1(\Bk)$ is a homogeneous function of $\Bk$.
%Then the quadratic quasiconvex function associated with $\BT$ is $f(\BP)=\BP\cdot\BT\BP$, having
%the quasiconvexity property that the integral over all space of $f(\BE)$ is non-negative when $\BGG_1\BE=\BE$. 
The non-negativity of the operator $\BGG_1\BT\BGG_1\geq 0$ is only linked to the
highest derivatives in the functional one is minimizing. This is because
these notions of quasiconvexity arose in the context of understanding what could go wrong
when seeking minimizers of nonconvex functions in
the calculus of variations.
Without going much into the details,
%weak lower semicontinuity is ensured if the finest scale oscillations are penalized.
weak lower semicontinuity is what one needs to show existence of smooth solutions to equations, such as the nonlinear
elasticity equations where one needs to show that minimizers of the integral of $W(\Grad\Bu(\Bx))$, over a body represented by a region $\GO$ in the undeformed state, 
exist \cite{Ball:1977:CCE,Ball:1981:NLW}. Here $W$ is the elastic energy and $\Bu(\Bx)$ is
the position of a particle in the body having coordinates $\Bx\in\GO$ in the undeformed state. 
Rather than there being a minimizer there may be a sequence
of highly oscillatory functions $\Bu=\Bu_i(\Bx)$, $i=1,2,3\ldots,$ producing ever lower energies that cannot be achieved with smooth functions $\Bu(\Bx)$.
If this happens one says that the integral as a function of $\Bu$ is not weakly lower semicontinuous.
Quasiconvexity safeguards against this by ensuring that the finest scale oscillations have an energy penalization.
When one has quadratic functions  $W(\Grad\Bu,\Grad\Grad\Bu,\ldots, \Grad^m\Bu)$ it is typically only the dependence of $W$ on
the highest derivatives $\Grad^m\Bu$ (which dominate when $\Bu$ is highly oscillatory) that is important to
determining weak lower semicontinuity and hence the existence of a minimizer.
One needs to show that $W(\Grad\Bu,\Grad\Bu,\ldots, \Grad^m\Bu)$ is quasiconvex with respect to $\Grad^m\Bu$ when one replaces all the other
arguments of $W$ by fixed constants \cite{Meyers:1965:ASA}. If $W$ is a quadratic function and a smooth minimizer exists,
it satisfies the $m$-th order gradient elasticity equations discussed in Section 3 of
Part IV \cite{Milton:2020:UPLIV}: hence the connection to
$\BGG_1(\Bk)$, with $\Grad^m\Bu$ being associated with the terms of order $\Bk^{2m}$ in $\BGG_1(\Bk)$. For bounding the
spectrum of $\BA$ we need $Q^*$-convexity rather than quasiconvexity if  $\BGG_1(\Bk)$ is not  a homogeneous function in $\Bk$.  For the Schr{\"o}dinger equation
some  $Q^*$-convex operators have been identified (see Sections 13.6 and
13.7 of \cite{Milton:2016:ETC}), but not yet applied to bounding spectrums.

Combining the equations in \eq{t.1} gives
\beq \BA\geq \BGG_1\BT\BGG_1+a^-\BGG_1 \geq a^-\BGG_1. \eeq{t.2}
Then clearly $a^-$ is a lower bound on the spectrum of $\BA= \BGG_1\BB\BGG_1$ in the space $\CE$. Alternatively, for the same  or another $\BT$
satisfying $\BGG_1\BT\BGG_1\geq 0$, one can look for a constant $a^+$ such that
\beq a^+\BI-\BB(\Bx)\geq \BT(\Bx)\quad\text{for all }\Bx, \eeq{t.2a}
and we obtain
\beq a^+\BGG_1-\BA(\Bx)\geq \BGG_1\BT\BGG_1\geq 0, \eeq{t.2b}
thus implying that $a^+$ is an upper bound on the spectrum of $\BA$. From the identity \eq{1.15a} we see that when $\BB$ is a finite dimensional matrix,
bounds on the spectrum of 
\beq \widetilde{\BA}=\BGG_2\widetilde{\BB}\BGG_2,\quad\text{where  }
\widetilde{\BB}=\BI/z_0-(z_0\BI-\BB)^{-1},
\eeq{t.2ba}
also allow us to bound the values of $z_0$ for which
$\BR-\BL^{-1}$ has a null space. Note that $z_0^2\widetilde{\BB}$ approaches $-\BB$ in
the limit as $z_0\to\infty$. 
To bound the spectrum of $\widetilde{\BA}$ we seek a $\widetilde{\BT}(\Bx)$ and constant $\widetilde{a}^-$
such that
\beq \widetilde{\BB}(\Bx)\geq  \widetilde{\BT}(\Bx)+ \widetilde{a}^-\BI\quad\text{for all }\Bx, \quad \BGG_2 \widetilde{\BT}\BGG_2\geq 0,
\eeq{t.1alt}
and then $\widetilde{a}^-$ is a lower bound on the spectrum of $\widetilde{\BA}$ (where this spectrum itself depends on $z_0$). Bounds analogous
to \eq{t.2b} can clearly also be obtained. When $\BB(\Bx)=\chi(\Bx)\BI$ for some characteristic function then, as observed following \eq{1.15a},
$\BGG_2\BB(\Bx)\BGG_2$ will have exactly the same spectrum as $\BI-\BA$, even though $\BGG_1$ and $\BGG_2$ project onto different spaces.

One of the most important classes of $\BT(\Bx)=\BT_{nl}(\Bx)$, not necessarily Hermitian, are those having the property that $\BGG_1\BT_{nl}\BGG_1=0$.
The associated quadratic form is then what is known as a null-Lagrangian, so we call them null-$\BT$ operators. However, it should
be remembered that the non-Hermitian part of $\BT_{nl}(\Bx)$ gets lost when considering the quadratic form --- the quadratic form could even
be zero if  $\BT_{nl}(\Bx)$ is anti-Hermitian. A simple example is for electrical conductivity with $\BGG_1(\Bk)=\Bk\otimes\Bk/k^2$
where one may take  $\BT_{nl}(\Bx)$ to be any antisymmetric matrix valued field with $\Div\BT_{nl}=0$.
If $\BGG_1\BE=\BE$, then automatically
\beq \BGG_1\BT_{nl}\BE=\BGG_1\BT_{nl}\BGG_1\BE=0. \eeq{nl1}
In other words, we are free to subtract $\BT_{nl}$ (or any multiple of it) from  $\BL(\Bx)$ without disturbing the solution $\BE(\Bx)$, and with $\BJ(\Bx)$
being replaced by $\BJ-\BT_{nl}\BE$. We can shift  $\BL(\Bx)$ in this way as we please. These $\BT_{nl}(\Bx)$ allow one to establish equivalence classes
between problems taking the form \eq{ad1}. Of course, if one is  finding the spectral bounds according to the prescription
just outlined, then the non-Hermitian part of $\BT_{nl}(\Bx)$ is irrelevant, and we may as well assume that $\BT_{nl}(\Bx)$ is Hermitian. Then one
can recover it from the associated null-Lagrangian. In this case we call $\BT_{nl}(\Bx)$ a null-Lagrangian.
To generate suitable null-$\BT$ operators in three dimensions, for equations where
$\BGG_1(\Bk)$ has some diagonal block entries of the form $\Bk\otimes\Bk/k^2$,
one uses the result that if $\BU(\Bx)$ is a antisymmetric $3\times 3$ matrix valued field with $\Div\BU=0$,
and $\Be(\Bx)$ is a three component curl free field, then $\Bj=\BU\Be$ satisfies $\Div\Bj=0$. This is easily seen if we write $\Be=\Grad \Gf$, giving
\beq \Div\Bj=\Div\BU(\Bx)\Grad\Gf=(\Div\BU)\cdot\Grad\Gf+\Tr(\BU\Grad\Grad\Gf)=0,
\eeq{nl2}
in which $\Tr(\BU\Grad\Grad\Gf)$ vanishes because $\BU$ is antisymmetric while $\Grad\Grad\Gf$ is symmetric. Equivalently, one may write $\BU=\BGn(\Bu)$
where the antisymmetric matrix $\BGn(\Bu)$ has the property that $\BGn(\Bu)\Be=\Bu\otimes\Bv$.
Then  $\Div\BU$ is zero if $\Curl\Bu=0$, and \eq{nl2} reflects the fact
that the cross product of two curl free fields is divergence free. 
Here $\Gf$ could be any potential, or linear
combination of potential components, in the equations that are being studied. The same holds true in two dimensions, but then the condition
that $\Div\BU=0$ forces the antisymmetric $2\times 2$ matrix $\BU$ to be constant, and thus proportional to
\beq \BR_\perp=\bpm 0 & 1 \\ -1 &0 \epm,
\eeq{nl3}
the matrix for a $90^\circ$ rotation. 
Additionally in two dimensions, $\BR_\perp$ acting on a divergence free field $\Bj$ produces a curl free field $\BR_\perp\Bj$. Through these
observations one can generate a multitude of null-$\BT$ operators
associated with a given operator $\BGG_1$.

As an application of the power of  null Lagrangians in proving uniqueness of solutions, one may consider the elasticity equations, when the
elasticity tensor $\BCC(\Bx)$ is bounded and coercive, i.e., on the space of symmetric matrices the inequality
\beq \Gb^+\BI \geq \BCC(\Bx) \geq \Gb^+\BI \eeq{bnc1}
holds for some constants $\Gb^+>\Gb^->0$. This implies that a unique solution for the strain
$\BGve=[\Grad\Bu+(\Grad\Bu)]$ exists, but that does that uniquely determine $\Bu$? Korn's inequality shows that it does,
but a simpler approach \cite{Kondratiev:1988:BVP} is to introduce the null-Lagrangian associated with a fourth order tensor $\CT$,  whose action on a matrix $\BP$ is given by
\beq \BCT\BP=\BI\Tr(\BP)-\BP^T. \eeq{bnc2}
Adding $\Ge\BCT$, with $\Ge>0$, to $\BCC(\Bx)$ gives an equivalent problem that breaks the degeneracy: for small enough $\Ge$, $\BCC(\Bx)+\Ge\BCT$ is bounded
and coercive on the space of all matrix valued fields, not just the symmetric matrix valued fields (see, for example, Section 6.4 of \cite{Milton:2002:TOC}). So $\Grad\Bu$, and hence $\Bu$, is uniquely determined.
In the context of minimizing sequences of fields, Bhattacharya \cite{Bhattacharya:1991:KIS} has used $\BCT$ to bound the fluctuations in the antisymmetric part of $\Bu$
in terms of the fluctuations of the symmetric part of $\Grad\Bu$. 

For a fixed $\BGG_1$ the associated set $\CS$ of Hermitian $Q^*$-convex operators is a convex set, since if $\BGG_1\BT_1\BGG_1\geq 0$
and $\BGG_1\BT_2\BGG_1\geq 0$ then clearly  $\BGG_1[(\BT_1+\BT_2)/2]\BGG_1\geq 0$. The set is invariant with respect to additions
or subtractions of any null-Lagrangians.  If we focus, for simplicity, on $Q^*$-convex operators that do not depend on $\Bx$, then it makes sense to
look for the extreme points of $\CS$, modulo additions or subtractions of null-Lagrangians. These extremal $\BT=\BT_e$ have the property
that they lose their $Q^*$-convexity whenever any
$Q^*$-convex $\BT$ that is not a null-Lagrangian is subtracted from it,
as portrayed in \fig{ext}.
Since
the inequalities
\beq  \BB(\Bx)\geq \BT_1 +a_1^-\BI \quad\text{and}\quad \BB(\Bx)\geq \BT_2 +a_2^-\BI\quad\text{imply}\quad
\BB(\Bx)\geq \tfrac{1}{2}(\BT_1+\BT_2) +\tfrac{1}{2}(a_1^-+a_2^-)\BI,
\eeq{tsp1}
we see that the best bounds on the spectrum will be generated by the extremal $\BT$. The characterization of the extremal $\BT$ is a challenging
problem. Interestingly, there is a connection with extremal polynomials: see \cite{Harutyunyan:2016:TCE} and references therein. 

\begin{figure}[!ht]
\centering
\includegraphics[width=0.95\textwidth]{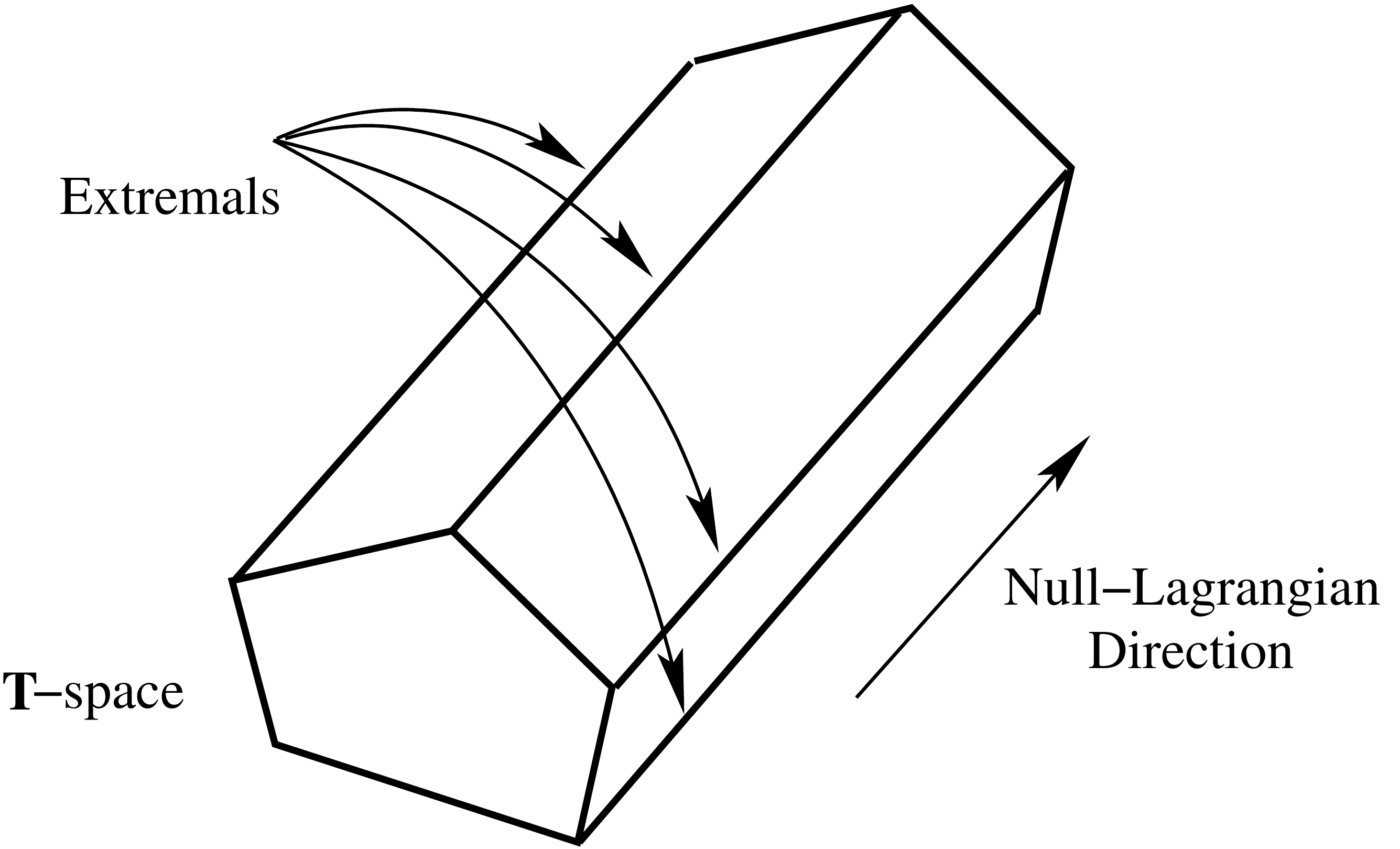}
\caption{A very schematic figure showing what is meant by extremal
  $Q^*$-convex operators $\BT$. This is a high dimensional space and
  there may be many directions that correspond to Null-Lagrangians.
  There is a cone of possible $\BT$, since if $\BT$ is $Q^*$-convex
  so is $\Gl\BT$ for all $\Gl>0$. The figure does not capture the
  cone, as that would make it at least four dimensional, but rather the figure
  can be viewed as a linear cross section through that cone.
  %For comparison
  %a positive semidefinite $3\times 3$ matrix $\BT\ne 0$ with eigenvalues
  %$\Gl_1, \Gl_2$ and $\Gl_3$ normalized with  $\Gl_1+\Gl_2+\Gl_3=1$,
  %is represented as a triangle being the intersection of the plane
  %$\Gl_1+\Gl_2+\Gl_3=1$ with the octant $\Gl_1>0$, $\Gl_2>0$ and $\Gl_3>0$
  %when one uses these eigenvalues 
}
\labfig{ext}
\end{figure}

Tighter spectral bounds can be obtained by
following important ideas of Murat and Tartar \cite{Murat:1978:CPC,Tartar:1979:CCA,Tartar:1979:ECH,Murat:1981:CPC,Murat:1985:CVH,Tartar:1985:EFC} and embedding
the problem in a coupled field equation setting where there are no couplings, but letting $\BT$ be chosen with couplings. 
Thus, given an integer $\ell\geq 1$, define
\beq
\mathbb{B}=\underbrace{\bpm \BB & 0 & \ldots & 0 & 0 \\
0  & \BB & \ldots & 0 & 0 \\
\vdots & \vdots & \ddots & \vdots &\vdots \\
0 & 0 & \ldots & \BB & 0 \\
0 & 0 & \ldots & 0 & \BB\epm}_{\ell\text{  blocks}}, \quad
\mathbb{G}_1=\underbrace{\bpm \BGG_1 & 0 & \ldots & 0 & 0 \\
0  & \BGG_1 & \ldots 0 & 0 \\
\vdots & \vdots & \ddots & \vdots \\
0 & 0 & \ldots & \BGG_1 & 0 \\
0 & 0 & \ldots & 0 & \BGG_1\epm}_{\ell\text{  blocks}}.
\eeq{t.3}
If for some tensor field $\mathbb{T}(\Bx)$ and constant $a^-$ one has
\beq \mathbb{B}\geq \mathbb{T}(\Bx)+a^-\BI\quad\text{for all }\Bx, \quad \mathbb{G}_1\mathbb{T}\mathbb{G}_1\geq 0, \eeq{t.3aa}
then it follows that
\beq  \bpm \BA & 0 & \ldots & 0 & 0 \\
0  & \BA & \ldots & 0 & 0 \\
\vdots & \vdots & \ddots & \vdots & \vdots\\
0 & 0 & \ldots & \BA & 0 \\
0 & 0 & \ldots & 0 & \BA\epm =\mathbb{G}_1\mathbb{B}\mathbb{G}_1
\geq \mathbb{G}_1\mathbb{T}\mathbb{G}_1+a^-\BI \geq a^-\BI.
\eeq{t.4}
As the first and last operators are both block diagonal we see that this new $a^-$ is also a lower bound on the spectrum of $\BA= \BGG_1\BB\BGG_1$ in the space $\CE$.
Of course if $\mathbb{T}$ is block diagonal then we gain nothing by this procedure, but the point is to take operators $\mathbb{T}$ where there are off diagonal blocks
that couple everything together.

Generally it is very difficult to find $\BT(\Bx)$ such that $\BGG_1\BT\BGG_1\geq 0$ (or $\mathbb{T}(\Bx)$ such that $\mathbb{G}_1\mathbb{T}\mathbb{G}_1\geq 0$ which
is essentially the same problem so we will not treat it separately). However there appear to be at least three routes. One approach, following
the ideas of Tartar and Murat \cite{Murat:1978:CPC,Tartar:1979:CCA,Tartar:1979:ECH,Murat:1981:CPC,Murat:1985:CVH,Tartar:1985:EFC}, is to look for
$\BT(\Bx)$ that are constant
and recognize $\BGG_1\BT\BGG_1\geq 0$ is an inequality in Fourier space with $\BGG_1\BT\BGG_1\geq 0$ acting locally in Fourier space.
Thus the inequality  holds if and only if
\beq \BGG_1(\Bk)\BT\BGG_1(\Bk)\geq 0\quad\text{for all}\quad\Bk.
\eeq{t.5}
This still leaves an abundance of choices of $\BT$. Taking $\BT$ independent of $\Bx$ can have some additional advantages too. In the field of composites,
with $\BL(\Bx)$ piecewise constant taking $N$ values corresponding
to $N$ phases it leads to bounds that do not depend upon the
microstructure of the composite, aside from possibly the volume fractions
occupied by the phases in the composite. While there are many examples of $\BT$ that have been worked out for a multitude of problems when
$\BGG_1(\Bk)$ is a homogeneous function of $\Bk$, the case where this is not so (as, for example, occurs in time harmonic wave equations,
time dependent diffusion of heat or particles and the Schr{\"o}dinger equations) is uncharted territory. I know only of examples associated
with the time harmonic Schr{\"o}dinger equation: see Sections 13.6 and 13.7 in \cite{Milton:2016:ETC}.

One general class of $\mathbb{T}(\Bx)$ (or $\BT$) has met with a lot of
success as it generates sharp bounds on the effective moduli
of composites corresponding to many obtained using the successful Hashin-Shtrikman variational principles \cite{Hashin:1962:VAT, Hashin:1963:VAT}, and sometimes
improves on them. This form of $\mathbb{T}$ \cite{Milton:1990:CSP} was motivated
by optimal bounds \cite{Kohn:1988:OBE, Avellaneda:1987:OBM} derived using
these  variational principles, and is given by:
\beq \mathbb{T}=\bpm \BL_0 & 0 & \ldots & 0 & 0 \\
0  & \BL_0 & \ldots & 0 & 0 \\
\vdots & \vdots & \ddots & \vdots & \vdots\\
0 & 0 & \ldots & \BL_0 & 0 \\
0 & 0 & \ldots & 0 & \BL_0 \epm-\Gv\BV\otimes\BV,\quad \BV=\bpm \Bv_1 \\ \Bv_2\\ \vdots \\ \Bv_{\ell-1} \\ \Bv_{\ell} \epm,
\eeq{t.5a}
where by choosing
\beq 1/\Gv=\min_\Bk\sum_{i=1}^\ell \Bv_i\cdot\BGG(\Bk)\Bv_i, \quad
\text{where}\quad \BGG(\Bk)=\BGG_1(\Bk)[\BGG_1(\Bk)\BL_0\BGG_1(\Bk)]^{-1}\BGG_1(\Bk),
\eeq{t.5b}
we ensure that $\mathbb{T}$ is $Q_*$-convex \cite{Milton:1990:CSP}.
Typically the ``reference medium'' $\BL_0$ is Hermitian and
positive definite, but it suffices for it to be  $Q_*$-convex. 

If $\BT$ is rotationally invariant then it will suffice to check this inequality
for one value of $\Bk$. There is some freedom in what one means by rotationally invariant. For example for three dimensional
conductivity with $\ell=3$ one may treat $\mathbb{T}(\Bx)$ as a fourth order tensor with three potentials that mix under rotations,
or as an array of 9 second order tensors, with three potentials that do not mix under rotations, and $\mathbb{T}(\Bx)$
may be rotationally invariant in one sense but not the other. In the first instance there are 3 real parameters that specify a rotationally invariant
Hermitian fourth order tensor (each associated with projections onto the three rotationally invariant subspaces of matrices proportional to $\BI$,
tracefree symmetric matrices and antisymmetric matrices), but three real and three complex numbers in a Hermitian matrix $\mathbb{T}(\Bx)$
containing 9 blocks each proportional to the identity matrix. In applications more success in producing tight bounds on the
effective properties of composites have  been obtained by looking for $\mathbb{T}(\Bx)$
that are fourth order tensors for conductivity \cite{Murat:1985:CVH, Tartar:1985:EFC, Avellaneda:1988:ECP} and eighth order tensors for elasticity
\cite{Milton:1990:BRT, Milton:1990:CSP}.

Two other routes can produce a $\BT(\Bx)$ that depend upon $\Bx$. This can be advantageous since $\BB(\Bx)$ depends on $\Bx$.  Given $\BGG_1(\Bk)$ one approach is to take an associated
$\BT(\Bx)$, and then to make a coordinate transformation in the underlying equations
to $\Bx'=\Bx'(\Bx)$ and obtain a $\BGG_1'(\Bk)$ and $\BT'(\Bx')$
in the new coordinates, that depends on $\Bx'$ even if $\BT(\Bx)$ was independent of $\Bx$. The second approach is to make a substitution.
For example, following \cite{Olver:1988:SNL} and Section V(C) of \cite{Milton:2019:NRF}, if some of the fields in $\CE$ derive from the derivatives
of any order of a scalar, vector, or tensor potential $\Bu$ then for any given
$\BZ(\Bx)$ and $\Bz(\Bx)$ one can try
substituting
\beq \widetilde{\Bu}(\Bx)=\BZ(\Bx)\Bu(\Bx)+\Bz(\Bx),
\eeq{t.6}
in a $Q^*$-convex quadratic form, involving $\widetilde{\Bu}$ and
its derivatives, to get a $Q^*$-convex quadratic form, and associated
 $Q^*$-convex operator $\BT$, involving $\Bu$ and its derivatives.
Here $\Bu$ might include
all potentials on the right hand side of the constitutive law.
Note that even if the original $Q^*$-convex quadratic form,
only involves say $\Grad\widetilde{\Bu}$ then
the new $Q^*$-convex quadratic form will involve both $\Grad\Bu$ and
$\Bu$. 
In this case $\BGG(\Bk)$ will transform to a $\widetilde{\BGG}(\Bk)$ that will not necessarily be a homogeneous function of $\Bk$ even if
$\BGG_1(\Bk)$ is a homogeneous function of $\Bk$. Correspondingly,
$\BT$ will transform to an associated $\widetilde{\BT}$
that generally will be a function of $\Bx$ even if $\BT$ was not.
We will not discuss these last two routes, but instead we refer the interested
reader to Section V in \cite{Milton:2019:NRF}.

We can also use $Q^*$-convex operators to bound the spectrum of the function $1/z_*(z_0)$ defined in Section \sect{31}. From \eq{r.3} we see this spectrum is contained in the spectrum of the operator $\BGG_1\BB\BGG_1$, so outer bounds on this spectrum immediately apply to
the spectrum of $1/z_*(z_0)$. Similarly, the spectrum of $z_*$ as a function of
$\und{z}_0=1/z_0$ is contained in the spectrum of $\BGG_2\und{\BB}\BGG_2$
and outer bounds on this spectrum immediately apply to the spectrum of $z_*$ as a function of
$\und{z}_0$.

The results in this Section are easily extended to non-Hermitian operators. For example, we can replace \eq{t.1} with
\beq e^{i\Gvt}\BA\geq a^-\BGG_1\quad\text{if}\quad
e^{i\Gvt}\BB(\Bx)\geq \BT(\Bx)+a^-\BI\quad\text{for all }\Bx\quad
\text{and}\quad\BGG_1\BT\BGG_1\geq 0, \eeq{t.100}
where the inequalities holds in the sense of quadratic forms, i.e.
they bound the Hermitian part of $e^{i\Gvt}\BA$ given bounds
on the Hermitian part of $e^{i\Gvt}\BB$. Similarly with $z_0=0$ so
that $\widetilde{\BB}(\Bx)$ is the inverse of $\BB(\Bx)$, the obvious
extension of \eq{t.100} implies bounds on the Hermitian part of
$e^{i\Gvt}\BGG_2\BB^{-1}\BGG_2$ given bounds on the Hermitian part of
$e^{i\Gvt}\BB^{-1}$.

%%%%%%%%%%%%%%%%%%%%%%%%%%%%%%%%%%%%%%%%%%%%%%%%%%%%%%%%%%%%%%%%%%%%%%%%%%%%%%%%%%%%%%%%%%%%%%%%%%%%%%%%%%%
\section{A remarkable identity between the resolvent of a non-Hermitian operator $\BA=\BGG_1\BB\BGG_1$
  and the inverse of an associated Hermitian operator}
\setcounter{equation}{0}
\labsect{36}
%%%%%%%%%%%%%%%%%%%%%%%%%%%%%%%%%%%%%%%%%%%%%%%%%%%%%%%%%%%%%%%%%%%%%%%%%%%%%%%%%%%%%%%%%%%%%%%%%%%%%%%%%%%%%%%%%%%%
Let us express $\BL$ as $\BL=\BL_1+\BL_2$ and consider the equation
\beq \BJ = (\BL_1+\BL_2)\BE-\Bs, \eeq{6.1}
which we rewrite in the two equivalent forms
\beqa \BE & = &\BL_1^{-1}\BJ-\BL_1^{-1}\BL_2\BE+\BL_1^{-1}\Bs,  \nonum
\BJ & = &\BL_1\BE+\BL_2[\BL_1^{-1}\BJ-\BL_1^{-1}\BL_2\BE+\BL_1^{-1}\Bs]-\Bs.
      \eeqa{6.3}
The first is easily seen to hold by substituting \eq{6.1} in it, and the second follows by substituting the first equation back in \eq{6.1}.   
We write these and the differential constraints as 
\beq \underbrace{\bpm \BE \\ \BJ \epm}_{\BJ^0}
=\underbrace{\bpm \BL_1^{-1} & -\BL_1^{-1}\BL_2 \\ \BL_2\BL_1^{-1} & \BL_1-\BL_2\BL_1^{-1}\BL_2\epm}_{\BL^0} \underbrace{\bpm \BJ \\ \BE \epm}_{\BE^0}
-\underbrace{\bpm -\BL_1^{-1}\Bs \\ \Bs-\BL_2\BL_1^{-1}\Bs \epm}_{\Bs^0},\quad \BGG^0_1=\bpm \BGG_2 & 0 \\ 0 & \BGG_1 \epm.
\eeq{6.4}
These manipulations are similar to the manipulations of Cherkaev and Gibiansky \cite{Cherkaev:1994:VPC} and the subsequent
manipulations in \cite{Milton:1990:CSP} that led to (4.4), (5.2), and (9.2)
in Part I \cite{Milton:2020:UPLI}, and which were generalized in \cite{Milton:2010:MVP} to include source terms. Now there is a close relation between $\BJ_0$ and $\BE_0$ and they need not be real. Clearly, we are back at a problem in the extended abstract theory of composites.
It so happens that the constitutive law \eq{6.1} implies this very close relation between $\BJ^0$ and $\BE^0$.
This equation has the solution
\beq \BE^0=[z^0\BI-\BGG^0_1\BB^0]^{-1}\BGG^0_1\Bs,\quad \text{where}\quad \BB^0=z^0\BI-\BL^0, \eeq{6.5}
implying
\beqa \BE & = &\bpm 0 & \BI \epm[z^0\BI-\BGG^0_1\BB^0]^{-1}\BGG^0_1\bpm -\BL_1^{-1}\Bs \\ \Bs-\BL_2\BL_1^{-1}\Bs \epm \nonum
& = & \bpm 0 & \BI \epm\BH^0\bpm -\BI\\ \BL_1-\BL_2\epm\BL_1^{-1}\Bs\quad\text{with}\quad\BH^0
=(\BGG^0_1\BL^0\BGG^0_1)^{-1}=[z^0\BI-\BGG^0_1\BB^0]^{-1}\BGG^0_1.
\eeqa{6.6}
Hence we arrive at the remarkable identity:
\beqa \BR & = &(z_0\BI-\BGG_1\BB\BGG_1)^{-1}=\BR_0-(\BGG_1-\BI)/z_0=[z_0\BI-\BGG_1\BB]^{-1}\BGG_1-(\BGG_1-\BI)/z_0 \nonum
          & = & \bpm 0 & \BI \epm\BH^0\bpm -\BI\\ \BL_1-\BL_2\epm\BL_1^{-1}-(\BGG_1-\BI)/z_0,
\eeqa{6.7}
that holds for any real or complex $\BB$ and any real or complex $z^0$ (not
to be confused with $z_0$)
with $\BB^0=z^0\BI-\BL^0$ giving $\BH^0$ as in \eq{6.6}, where $\BL^0$ is defined by
\eq{6.4}. 
In particular, if we take $\BL_1$ as the Hermitian part of $\BL$ and
$\BL_2$ as the anti-Hermitian part,
\beq \BL_1=(\BL+\BL^\dagger)/2,\quad \BL_2=(\BL-\BL^\dagger)/2,
\eeq{6.8}
then $\BL^0$ is a Hermitian operator. So we have an identity between the resolvent of a non-Hermitian operator and the inverse of an associated Hermitian operator.
Furthermore, and what is more significant,
$\BL^0$ will be positive definite if and only if $\BL_1$ is positive definite. 
Of course these results apply to matrices as well, not just operators.

The simplest example is when $\BGG_1=\BI$ and $\BB=(z_0-z)\BI$, where $z=z_1+iz_2$ is complex. Then
\beq \BGG^0_1\BB^0\BGG^0_1=\BGG^0_1(z^0\BI-\BL^0)\BGG^0_1=\bpm 0 & 0 \\ 0 & (z_0-z_1-z_2^2/z_1)\BI \epm, \eeq{6.9}
so that
\beq [z^0\BI-\BGG^0_1\BB^0\BGG^0_1]^{-1}=\bpm (z^0)^{-1}\BI & 0 \\ 0 & (z_1+z_2^2/z_1)^{-1}\BI \epm. \eeq{6.10}
Hence the right hand of \eq{6.7} evaluates to
\beq \bpm 0 & \BI \epm\bpm (z^0)^{-1}\BI & 0 \\ 0 & (z_1+z_2^2/z_1)^{-1}\BI \epm\bpm 0 & 0 \\ 0 & \BI \epm\bpm -\BI\\ z_1-iz_2\epm z_1^{-1}
= \frac{(z_1-iz_2)\BI}{z_1^2+z_2^2}=\frac{\BI}{z}=\BR,
\eeq{6.11}
in agreement with \eq{6.7}.

%%%%%%%%%%%%%%%%%%%%%%%%%%%
\section{A novel Stieltjes function integral representation
  for the resolvent of a non-Hermitian operator $\BA=\BGG_1\BB\BGG_1$}
 \setcounter{equation}{0}
\labsect{37}
%%%%%%%%%%%%%%%%%%%%%%%%%%%%%%%%%%%%%%%%%%%%%%%%%%%%%%%%%%%%%%%%%%%%%%%%%%%%%%%%%%%%%%%%%%%%%%%%%%%%%%%%%%%%%%%%%%%%
Here we obtain Stieltjes type
  integral representations for the resolvent in the case where $\BB$ is
  non-selfadjoint but there exists an angle $\vartheta$
  such that $c-[e^{i\vartheta}\BB+e^{-i\vartheta}\BB^\dagger]$ is positive definite
  (and coercive) for some constant $c$. The integral representation holds
  in the half plane $\Real(e^{i\vartheta}z_0)>c$. We just treat the case where
  $\vartheta=0$ as the extension to the case where  $\vartheta \ne 0$ is
  clear. 

  It is obviously best to keep $z_0$ complex rather than splitting it into its real and imaginary parts. To do this we take
\beq \BL_1=\BL_1(z_0)=z_0\BI-\tfrac{1}{2}(\BB+\BB^\dagger),\quad \BL_2=\tfrac{1}{2}(\BB^\dagger-\BB). \eeq{6.11aa}
%so that $\BL_1$ is no longer Hermitian when $z_0$ is complex. 
%\eq{6.4} as 
%\beq \underbrace{\bpm -i\BE \\ \BJ \epm}_{\underline{\BJ}^0}
%=\underbrace{\bpm -\BL_1^{-1} & i\BL_1^{-1}\BL_2 \\ -i\BL_2\BL_1^{-1} & \BL_1-\BL_2\BL_1^{-1}\BL_2\epm}_{\underline{\BL}^0} \underbrace{\bpm i\BJ \\ \BE \epm}_{\underline{\BE}^0}
%-\underbrace{\bpm -\BL_1^{-1}\Bs \\ \Bs-\BL_2\BL_1^{-1}\Bs \epm}_{\underlin{\Bs}^0},\quad \BGG^0_1=\bpm \BGG_2 & 0 \\ 0 & \BGG_1 \epm.
%\eeq{6.20}
Setting
\beq
\BZ(z_0)=\BL_1^{-1}=[z_0\BI-\tfrac{1}{2}(\BB+\BB^\dagger)]^{-1}=
\BZ'(z_0)+i\BZ''(z_0),
\eeq{6.21}
where $\BZ'(z_0)$ and $\BZ''(z_0)$ are the real
and imaginary parts of $\BZ(z_0)$, we see that $\BZ'(z_0)$ and $\BZ''(z_0)$
are Hermitian, and $\BZ'(z_0)$ is positive definite if
\beq \Real(z_0)\BI > \tfrac{1}{2}(\BB+\BB^\dagger). \eeq{6.22}
This is an extension of the result that the inverse of a matrix
$\BA=\BA_h+\BA_a$ with positive definite Hermitian part $\BA_h$ and
anti-Hermitian part $\BA_a$ has a positive definite Hermitian part.
To establish this we write
\beq \BA=\BA_h^{-1/2}[\BI-i(i\BA_h^{-1/2}\BA_a\BA_h^{-1/2})]^{-1}\BA_h^{-1/2},
\eeq{6.21aa}
and then diagonalize the Hermitian matrix $i\BA_h^{-1/2}\BA_a\BA_h^{-1/2}$
to calculate the inverse.
The Hermitian part of $\BL_0(z_0)$ is
\beqa \tfrac{1}{2}[\BL^0+(\BL^0)^\dagger]
& = & \bpm \BZ' & -\BZ'\BL_2 \\ \BL_2\BZ' &
\tfrac{1}{2}[\Real(z_0)\BI-\tfrac{1}{2}(\BB+\BB^\dagger)]
+\BL_2\BZ'\BL_2^\dagger\epm \nonum
& = & \bpm 0 & 0 \\ 0 & \Real(z_0)\BI-\tfrac{1}{2}(\BB+\BB^\dagger)
\epm
+\bpm \BI \\ -\BL_2^\dagger \epm\BZ'\bpm \BI  & -\BL_2 \epm,
\eeqa{6.23}
and this is clearly positive definite if \eq{6.22} holds.
Furthermore, the real and imaginary parts of $\BL^0(z_0)$ are each
Hermitian by themselves. Let $c$
be a real value of $z_0$ such that \eq{6.22} holds
(and $c-\tfrac{1}{2}(\BB+\BB^\dagger)$ is coercive), and define
$w_0=z_0-c$. Then we have
\beq  \BL_1=z_0\BI-\tfrac{1}{2}(\BB+\BB^\dagger)=
w_0\BI+\BM_0,\quad\text{with}\quad
\BM_0(\Bx)=c\BI -\tfrac{1}{2}[\BB(\Bx)+\BB^\dagger(\Bx)]> 0.
\eeq{6.23g}
So $\BH^0(w_0)$ is Hermitian when
$w_0$ is real and positive, and the
Hermitian part of  $\BH^0(w_0)$ is positive definite for all $w_0$
in the right hand plane. Additionally, as $w_0\to\infty$,
we have
\beq \BH^0(w_0)=w_0\BH_1+\mathcal{O}(1),\quad
\BH_1= \bpm \BGG_2 & 0 \\ 0 & 0 \epm.
%\BH^0(w_0)=w_0\BH_1+\BH_2+\mathcal{O}(1/w_0),\quad
%\BH_1= \bpm \BGG_2 & 0 \\ 0 & 0 \epm,\quad
%\BH_2= \bpm \BGG_2\BM_0\BGG_2 & 0 \\ 0 & 0 \epm.
\eeq{6.23h}
These properties are reminiscent of the
complex conductivity tensor $\BGs$ as a function of $-i\Go$ where $\Go$
is the frequency. The associated permittivity $\BGve=i\BGs/\Go$ is then
a Stieltjes function of $-\Go^2$: see, for example, \cite{Milton:1997:FFR}. 
Analogously, with $-i\Go$ replaced with $w_0$ and
setting $v=w_0^2$ we have that $\BH^0/w_0$ is a
a operator valued Stieltjes function of $v$. Equivalently, $\BH^0(w_0)/w_0$ has 
the representation formula:
\beq \BH^0(w_0)/w_0=\BF(v)=\BH_1+\int_0^\infty\frac{d\BGm(\Gl)}{v+\Gl}=
\BH_1+\int_0^\infty\frac{d\BGm(\Gl)}{w_0^2+\Gl},
\eeq{6.24}
where $\BGm(\Gl)$ is a positive semidefinite Hermitian valued measure.
This measure is given by the Stieltjes inversion formula:
for all $\Gl_2>\Gl_1\geq 0$,
\beq \tfrac{1}{2}[\BGm(\Gl_1)+\BGm(\Gl_2)]+\BGm((\Gl_1,\Gl_2))
=-\pi^{-1}\lim_{\Ge\downarrow 0}\int_{\Gl_1}^{\Gl_2}
\Imag[\BF(-\Gl+i\Ge)]\,d\Gl,
\eeq{6.25}
where $\Imag$ denotes the imaginary part
In summary, by substituting \eq{6.24} back in \eq{6.7} we obtain an
integral representation for $\BR$. We have established the following:
\bigskip 

\noindent {\bf Theorem 1}

\medskip
{\it   Given an operator $\BB$ and a real constant $c$ such that $c-(\BB+\BB^\dagger)$ is coercive, let $\BL_1(z_0)$, $\BL_2$, and $\BL^0(z_0)$ be as given by \eq{6.11aa}
  and \eq{6.4}. Then we have the resolvent identity
  \beq (z_0\BI-\BGG_1\BB\BGG_1)^{-1}
 =  \bpm 0 & \BI \epm\BH^0(w_0)\bpm -\BI\\ \BL_1-\BL_2\epm\BL_1^{-1}-(\BGG_1-\BI)/z_0,
\eeq{The1}
in which $w_0=z_0-c$ and 
\beq \BH^0(z_0-c)=(\BGG^0_1\BL^0(z_0)\BGG^0_1)^{-1}, \quad
\BGG^0_1=\bpm \BGG_2 & 0 \\ 0 & \BGG_1 \epm.
\eeq{The2}
Furthermore, $\BH^0(w_0)$ has the integral representation \eq{6.24}
in terms of
\beq \BH_1= \bpm \BGG_2 & 0 \\ 0 & 0 \epm, \eeq{The3}
and the positive semidefinite operator valued measure
$\BGm(\Gl)$ given by \eq{6.25}.}

\medskip
\noindent
We emphasize that the integral representation only holds for $z_0$
in the half plane $\Real(z_0)>c$. 

There are other families of non-selfadjoint operators for which the 
resolvents have integral representations. The simplest is for bounded operators
where the real and imaginary parts are each selfadjoint with a real part 
that is coercive (as for the just mentioned
complex conductivity tensor $\BGs$ as a function of $i\Go$).
For dissipative operators, which (modulo multiplication by a complex
number) have a positive semi-definite imaginary part, one can embed the
Hilbert space on which $\BA$ acts
in a larger Hilbert space and find a Hermitian operator $\BH$
such that $<f(\BH)\BP,\BQ>=<f(\BA)\BP,\BQ>$ for all $\BP$ and $\BQ$ in the Hilbert space
where $\BA$ acts, where $<\,,\,>$ denotes the norm in this Hilbert space
\cite{SzNagy:1970:HAO}. The spectral theory for $\BH$ then
allows one to compute $f(\BH)$ for analytic functions $f$, and gives
a Nevanlinna-Herglotz representation integral representation for the
resolvent associated with $\BH$. This result was anticipated by Liv\v{s}ic
in his construction of a characteristic function of a dissipative operator.
These and further mathematical developments
in the area have been summarized by
Kuzheel' \cite{Kuzheel:1993:ECC} and Pavlov \cite{Pavlov:1996:SAD}.
From the physics perspective, an excellent treatment of embedding
dissipative problems in a larger Hilbert space in which energy is
conserved, and also allowing for dispersion (frequency dependent moduli)
has been given by Figotin and Schenker \cite{Figotin:2005:STT}
(see also \cite{Tip:1998:LAD}).
As they point out, one can think of the additional
fields as corresponding to a system with an infinite number of
``hidden variables'' that may also be called a heat bath.
While one can easily go from a conservative system
with an infinite number of
hidden variables to a dissipative system, they show the reverse is true
too.

By contrast, our analysis does not correspond to introducing an
infinite number of ``hidden variables'' and applies simply when one
has a finite dimensional Hilbert space ($n$-dimensional vector space)
in which $\BA$ and $\BB$ are non-Hermitian $n\times n$ matrices.
In that case, $\BH^0/w_0$ is a $2n\times 2n$ matrix valued
Stieltjes function of $v=w_0^2=(z_0-c)^2$. The measure entering the
spectral representation will not be discrete, in contrast to the 
usual spectral representations of Hermitian matrices. This measure is then a 
positive semidefinite Hermitian  $2n\times 2n$ matrix valued
continuous measure. The ``heat bath'' approach corresponds to embedding in a
problem with Hermitian  $(n+m)\times (n+m)$ matrices, as $m$ (which can be thought of
as the number of oscillators) approaches infinity. A beautiful physical demonstration
of the energy absorbing properties of a system of undamped oscillators (pendulums)
is in \cite{Akay:2005:EVA}. 

%``hidden variables''
%The Cayley transform $\BC=(\BI-\BA)(\BI+\BA)^{-1}$
%maps dissipative operators to contractive ones, with operator norm less than 1, and there is an analogous result for contractive operators:
%the Bela Sz.-Nagy dilation theorem
%(Sz.-Nagy and Foia{\c{s} \cite{Nagy:1970:HAO}) embeds the Hilbert space on which $\BC$ acts in a larger Hilbert space and finds a unitary operator
%$\BU$ (satisfying $\BU^\dagger\BU=\BU\BU^\dagger=\BI$)
%such that $f(\BU)\BP=f(\BC)\BP$ for all $\BP$ in the Hilbert space where $\BC$ acts. However, as pointed out to me by
%Kirill Cherednichenko of Bath University, there is no easy way of passing between the results of Gohberg and Krein and Sz.-Nagy and Foia{\c{s}.
The approach we take here also has some similarities with Liv\v{s}ic's compression of resolvents: see \cite{Howland:1975:LMP, Livsic:1973:OOW} and references therein. 
%%%%%%%%%%%%%%%%%%%%%%%%%%%%%%%%%%%%%%%%%%%%%%%%%%%%%%%%%%%%%%%%%%%%%%%%%
\section*{Acknowledgements}
GWM thanks the National Science Foundation for support through grant DMS-1814854.
Mihai Putinar is thanked for drawing the author's attention to works on
the embedding of non-selfadjoint operator problems in selfadjoint ones.
Alex Figotin is thanked for helpful correspondence, and
for providing pertinent references.
%%%%%%%%%%%%%%%%%%%%%%%%%%%%%%%%%%%%%%%%%%%%%%%%%%%%%%%%%%%%%%%%%%%%%%%%%%%%%%%%%%%%%%%%%%
%\bibliographystyle{plain}
%\bibliography{/home/milton/tcbook,/home/milton/newref}
%%%%%%%%%%%%%%%%%%%%%%%%%%%%%%%%%%%%%%%%%%%%%%%%%%%%%%%%%%%%%%%%%%%%%%%%%
\ifx \bblindex \undefined \def \bblindex #1{} \fi\ifx \bbljournal \undefined
  \def \bbljournal #1{{\em #1}\index{#1@{\em #1}}} \fi\ifx \bblnumber
  \undefined \def \bblnumber #1{{\bf #1}} \fi\ifx \bblvolume \undefined \def
  \bblvolume #1{{\bf #1}} \fi\ifx \noopsort \undefined \def \noopsort #1{}
  \fi\ifx \bblindex \undefined \def \bblindex #1{} \fi\ifx \bbljournal
  \undefined \def \bbljournal #1{{\em #1}\index{#1@{\em #1}}} \fi\ifx
  \bblnumber \undefined \def \bblnumber #1{{\bf #1}} \fi\ifx \bblvolume
  \undefined \def \bblvolume #1{{\bf #1}} \fi\ifx \noopsort \undefined \def
  \noopsort #1{} \fi

\end{document}